\documentclass[conference]{IEEEtran}
\IEEEoverridecommandlockouts
\usepackage{cite}
\usepackage{amsmath,amssymb,amsfonts, url}
\usepackage{algorithmic}
\usepackage{graphicx}
\usepackage{textcomp}
\usepackage{xcolor}
\usepackage{booktabs}
\usepackage{multirow}
\usepackage{etoolbox}

\def\x / \def\L

\def\BibTeX{{\rm B\kern-.05em{\sc i\kern-.025em b}\kern-.08em
    T\kern-.1667em\lower.7ex\hbox{E}\kern-.125emX}}
\begin{document}

\def\mySystem{NouveauVoice}
\title{\mySystem: Generating Novel Pseudo Speakers for Voice Anonymization}


\author{
  \IEEEauthorblockN{Meiying Melissa Chen, Anastasia Kuznetsova, Zhenyu Wang and Zhiyao Duan}
  \IEEEauthorblockA{
    \textit{Department of Electrical and Computer Engineering, University of Rochester} \\
    Rochester NY, USA \\
    meiying.chen@rochester.edu
  }
}

\maketitle

\begin{abstract}
Advanced neural technologies in speech synthesis and voice conversion (VC) have introduced severe risks to personal privacy, necessitating robust Speaker Anonymization Systems (SAS). Existing SAS approaches modify voice characteristics in the hand-crafted feature space or speaker embedding space, often struggling to provide sufficient identity variance across generated voices. In this paper, we propose NouveauVoice, a novel pseudo-speaker generation framework based on a Hierarchical Deep Variational Autoencoder (NVAE). Integrated as a standalone plug-in module on top of state-of-the-art architectures (FACodec and CosyVoice2), our approach leverages tractable sampling and the Evidence Lower Bound (ELBO) objective to synthesize highly expressive pseudo-speaker embeddings with significantly enhanced speaker diversity. Evaluating our framework under a protocol similar to the VoicePrivacy Challenge alongside Maximum Mean Discrepancy (MMD) analysis, we demonstrate that NouveauVoice achieves strong identity concealment, yielding an Equal Error Rate (EER) exceeding 38\% against an automatic speaker verification attacker model. Our system shows a reasonable trade-off between strict anonymity, rich pseudo-speaker diversity, and downstream speech utility, such as intelligibility and emotional expressiveness.
\end{abstract}

\vspace{-3pt}
\section{Introduction}
\label{sec:int}
Advanced neural technologies, while providing great convenience in speech and audio production pipelines, pose significant challenges to personal privacy. Modern text-to-speech (TTS) systems \cite{ju2024naturalspeech, du2024cosyvoice, wang2025sparktts, chen2022controlvc} with voice conversion (VC) capabilities, large language models (LLMs) based on discrete speech tokenizers \cite{yang2025llms, huang2026kanade, Aihara2026,mousavi2025discrete} and numerous standalone VC systems \cite{qian2019autovc, huang2022sslvc, zhang2025vevo, chen2022controlvc, kameoka2018stargan, hussain2023ace} can accurately clone a target speaker’s identity with arbitrary content, making them susceptible to misuse. Furthermore, identity leakage from insufficiently concealed speech may allow a perpetrator to reveal sensitive information, e.g., in HIPAA-protected settings where specific attributes, such as accent or pathological conditions, can compromise anonymity. Consequently, speaker anonymization systems (SAS) remain a critical area of active research.

We distinguish between two primary architectures for SAS. First, digital signal processing (DSP) based systems \cite{patino21_interspeech, Tavi2026f0} are straightforward to implement by leveraging shifts in frequency formants, pitch, and other inherent acoustic signal characteristics. However, these methods remain highly vulnerable to sophisticated adversarial attacks, such as automatic speaker verification (ASV) systems retrained on anonymized data to uncover original speaker identities \cite{miao2023householder}.

Second, more advanced Deep Neural Network (DNN) based SAS frameworks comprise the recent baselines for the VoicePrivacy Challenge \cite{vpc2024, vpc2026}. These include Generative Adversarial Network (GAN) based pseudo-speaker generation (B3) and Neural Audio Codec (NAC) systems that utilize a pool of random prompts to condition an autoregressive decoder, modifying the acoustic tokens that encode speaker identity (B4) \cite{vpc2026}. More recently, Large Language Model (LLM) based SAS, such as StreamVoiceAnon \cite{kuzmin2026streamvoiceanon}, utilize acoustic features or speech tokens from the target voice as prompts. Notably, \cite{kuzmin2026streamvoiceanon} employs a pseudo-speaker selection strategy that extracts speaker embeddings from a reference prompt pool, averages them, and injects Gaussian noise to enhance privacy. While highly efficient for streaming configurations, such pooling methods do not explicitly guarantee rich speaker diversity or natural acoustic variability across the generated pseudo-identities.

The main principles for speaker anonymization that we focus on in this paper are: 
i) \textit{Speaker privacy protection}. Generated pseudo-speakers should be dissimilar to the original speaker; ii) \textit{Speaker diversity}. Pseudo-speakers should have unique speaker identity to maintain the diversity of anonymized speech across different speakers; iii) \textit{Distribution similarity.} The distribution of pseudo-speakers should be close to the original speaker distribution to maintain the naturalness of the original speech. These requirements are also concisely formulated in \cite{noe2020pseudonymization} and \cite{miao2023householder}.

Following these principles, we propose NouveauVoice, a novel approach for pseudo-speaker generation based on a Hierarchical Deep Variational Autoencoder (NVAE) \cite{vahdat2020nvae} that fits the DNN-based SAS paradigm. NouveauVoice is trained independently as a standalone plug-in module and validated across two state-of-the-art voice conversion backends: FACodec \cite{ju2024naturalspeech} and CosyVoice2 \cite{du2024cosyvoice}.
The training of the proposed NVAE is guided by the Evidence Lower Bound (ELBO) objective \cite{vahdat2020nvae} to enforce the distribution similarity to the original speaker. 

We conduct a comprehensive evaluation of NouveauVoice for pseudo-speaker generation following the evaluation strategy similar to Voice Privacy Challenge \cite{vpc2024, vpc2026} as well as speaker diversity evaluation via  Maximum Mean Discrepancy (MMD). 
The proposed system shows over ~38~\% Equal Error Rate (EER) against an ASV attacker model indicating strong anonymization capacity, while showing a reasonable trade-off between anonymity and intelligibility. 

The contributions of this paper are as follows:

\begin{itemize}
    \item The proposed approach is a convenient plug-in module that can turn existing VC systems into SAS;
    \item It offers fast and tractable generation of pseudo-speakers focusing on the speaker diversity requirement;
    \item Allows flexible control over anonymization strength.
\end{itemize}

\section{Method}
\label{sec:met}
Voice conversion (VC) is a generative speech processing framework that allows to change the personality of the source speaker to any desirable target identity. However, under privacy protection scenarios, the identity of neither source, nor target speaker may not be discovered to avoid malicious use of the speaker identities. Thus, we propose NouveauVoice, the hierarchical pseudo-speaker generator based on Nouveau Variational Auto Encoder (NVAE) \cite{vahdat2020nvae}, to model the speaker embedding distribution.

\subsection{NVAE Pseudo-Speaker Generator}
NVAE was first introduced by \cite{vahdat2020nvae} for high-quality image generation. 
Its encoder transforms the input feature vector $x$ into the latent vector $z$, which is partitioned into disjoint groups of latent sub-vectors from coarser levels to finer levels, $z = (z_1, \ldots, z_L)$, where $L$ is the total number of latent groups and each $z_l$ represents the latent vector for group $l$. The generative model $p_\theta(x| z)$ is parametrized by $\theta$ via deep neural decoder and is represented by a series of conditional products:
\begin{equation}
p_\theta(x, z) = p_\theta(x|z) \prod_{l=1}^{L} p_\theta(z_l|z_{<l}),
\end{equation}
where $p_\theta(z_1|z_{<l}) = p_\theta(z_1)$ forms the base case prior, $z_{<l} \equiv (z_1, \ldots, z_{l-1})$ represents all coarser-level latents, $p_\theta(x|z)$ is the observation model, and $p_\theta(z_l|z_{<l})$ is the conditional prior for $z_l$ given latents from coarser levels $z_{<l}$.

The encoder $q_\phi(z|x)=\prod_{l=1}^L q_\phi(z_l|z_{<l},x)$ (where $\phi$ denotes the encoder parameters) shares the autoregressive ordering of the prior. A deterministic pass extracts fearues from $x$, the groups are then sampled from coarse to fine, with each approximate posterior $q_\phi(z_l|z_{<l},x)$ conditioning on the coarser, already sampled groups $z_{<l} \equiv (z_{1}, \ldots, z_{l-1})$.

NVAE stabilizes training via spectral regularization  and residual cells with skip connections. The model optimizes the Evidence Lower Bound (ELBO):
\begin{equation}\label{eq:elbo}
	\begin{split}
&\mathcal{L}(\theta, \phi; x) = \mathbb{E}_{q_\phi(z|x)} \left[ \log p_\theta(x|z) \right] \\
\quad - & \sum_{l=1}^{L} \mathbb{E}_{q_\phi(z_{<l}|x)} \left[ D_{\text{KL}}(q_\phi(z_l|z_{<l},x) \| p_\theta(z_l|z_{<l})) \right],
\end{split}
\end{equation}
where the first term represents the reconstruction likelihood and the second term contains the KL divergence between the approximate posterior $q_\phi(z_l|z_{<l},x)$ and conditional prior $p_\theta(z_l|z_{<l})$ at each level $l$.

The expressivity of the model is supported by depth-wise convolutions increasing their receptive field with every layer, that helps to model long-range dependencies as well as fine-grained speaker nuances. Residual connections along with spectral regularization not only tame numerical instability of the VAE but also explicitly ensures the closeness of the posterior approximation to a real speaker distribution, which is one of the requirements for a good speaker anonymization system.

To accommodate speaker embeddings in the training data, we replace 2D convolutional layers with 1D equivalents, apply quantile-based normalization \cite{merad2023robust} to speaker embeddings, and introduce free-bits regularization \cite{kingma2016improved} with KL coefficient warmup \cite{higgins2017beta} to prevent posterior collapse.

\begin{figure*}[htbp]
    \centering
    \includegraphics[width=0.8 \linewidth]{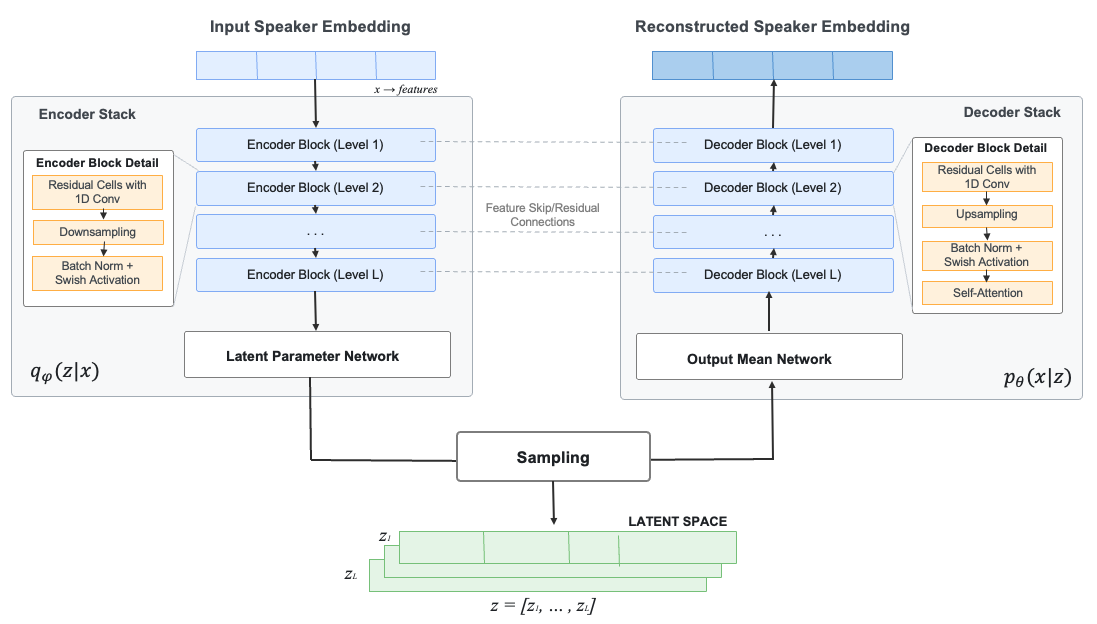}
    \caption{NouveauVoice generator overview. Left side shows the encoder structure, while right side depicts the decoder structure. Dashed lines between encoder and decoder blocks denote residual connections.}
    \label{fig:system}
\end{figure*}

\subsection{NouveauVoice Training and Inference}
Figure \ref{fig:system} shows NouveauVoice generator architecture and training scheme. During the training phase, speaker embeddings are passed to the NVAE multi-level CNN encoder stack consisting of 8 groups and each containing cells with two BN-Swish Conv1D layers (kernel size 3, stride 1 or 2) \cite{vahdat2020nvae}, where downsampling cells expand the receptive field progressively with depth.
Encoded hierarchical latent variables $(z_1 \dots z_L)$, which is jointly parameterized by encoder and decoder features through residual connections, and passed to the decoder for reconstruction.
The ELBO objective (\ref{eq:elbo}) is calculated between the input speaker embeddings and their reconstruction.

During inference, latent codes $z_1, \dots, z_L$ are sampled sequentially through the trained hierarchical structure. At the highest level, a latent code $z_1$ is drawn from the unconditional prior $p_{\theta}(z_1)$. Then each subsequent latent code is sampled conditionally based on the previously generated groups, represented as $p_{\theta}(z_l | z_{<l})$. Finally, the complete latent representation $z = (z_1, \dots, z_L)$ is passed through a decoder to generate the pseudo speaker embedding in the original speaker embedding space.

\section{Experimental Setup}
\label{sec:exp}
In this work, we conduct two experiments to evaluate the performance and structural utility of the NVAE framework for pseudo-speaker generation. The first experiment evaluates the robustness of the generated speaker embeddings for anonymization when integrated into different VC systems. The second experiment investigates how the hierarchical architecture of the NVAE can be leveraged to provide configurable levels of privacy protection.

\subsection{Experiment 1}
To demonstrate the cross-system generalizability and privacy-utility trade-off of NouveauVoice, our first experimental configuration leverages two distinct, state-of-the-art voice conversion systems as synthesis engines: FACodec \cite{ju2024naturalspeech} and CosyVoice2 \cite{du2024cosyvoice}.

FACodec \cite{ju2024naturalspeech} is a factorized neural speech codec used for the inherent embedding setup. It explicitly decomposes the speech waveform into three distinct subspaces modeled by individual vector quantizers (VQ) \cite{vandenoord2017vqvae}: prosody, content, and acoustic detail. To prevent information leakage between these branches, FACodec utilizes gradient reversal layer (GRL) \cite{ganin2015grl} and an adversarial training framework. Additionally, a standalone timbre extractor is employed, which aids in retaining speaker-specific characteristics and ensuring clean latent space disentanglement. Due to this modular decomposition of speech components, FACodec possesses a strong zero-shot VC capability, allowing us to evaluate novel speaker embeddings without any data-specific fine-tuning. 
Further, we refer to this model as FACodec-NV.

In the training phase of NouveauVoice pseudo-speaker generator we use FACodec's speaker encoder to extract speaker embeddings from the \textit{train-clean-100} and \textit{train-clean-360} subsets of the LibriTTS \cite{zen2019libritts}. In total, the training set resulted in approximately 150k embeddings. During inference, discrete content tokens are extracted from the source speech via FACodec and concatenated with a novel pseudo-speaker embedding sampled from NouveauVoice generator. This combined representation is then fed into the FACodec's decoder for anonymization.



CosyVoice2 \cite{du2024cosyvoice} is a zero-shot text-to-speech (TTS) and voice conversion (VC) model that factorizes speech generation into three successive modules: a supervised semantic tokenizer, a unified text-speech language model, and a chunk-aware causal flow-matching decoder. To provide timbre information during the language model stage, the system employs a CAM++ speaker embedding \cite{wang2023campp}, which is pre-trained on a speaker verification task. 
We integrate our NVAE model into the CosyVoice2 architecture by replacing the original CAM++ speaker embedding with the NouveauVoice output. The training and inference processes for this integrated system, denoted as CosyVoice-NV follow the same procedures described for the FACodec-NV model. For our experiments, we utilize the official implementation and pre-trained checkpoints \cite{cosyvoicegithub} for both the main CosyVoice2 model and CAM++ submodule.


As a baseline for speaker embedding generation, we fit a Gaussian Mixture Model (GMM) \cite{reynolds2009gaussian, tacospawn} with $k=16$ components and diagonal covariance with per-dimension variance to the training set of speaker embeddings \cite{sklearn2011}. We draw 1000 samples from the fitted GMM and use them as speaker embeddings for speech synthesis. These generated samples were subsequently used as speaker embeddings for the FAcodec and Cosyvoice speaker module to synthesize audio waveforms, those models are denoted as FACodec-GMM and Cosyvoice-GMM respectively. 

\subsection{Experiment 2}
\label{sec:exp2}

To evaluate how different latent groups contribute to speaker identity, we designed a progressive replacement experiment. A speaker's voice is first encoded into 8 latent groups $(z_1,\dots,z_8)$ using the trained NVAE model, where earlier groups capture coarser, more global aspects of identity. To produce anonymized speech, we replace the first $N$ groups with samples drawn from the prior distribution (i.e., noise with no speaker-specific information), while keeping the remaining groups intact. We denote the resulting embedding as $\tilde{x}^{(N)}$. At $N = 0$, $\tilde{x}^{(0)}$, serves as a reconstruction baseline where all groups are preserved and the original identity is fully retained. As $N$ increases from 1 to 8, progressively more groups are replaced, stronger anonymization is achieved. At $N = 8$, $\tilde{x}^{(8)}$, all groups are replaced entirely, producing a fully anonymized embedding. This design allows us to study the contribution of each hierarchical level to speaker identity in a controlled manner, identifying the minimum number of groups that must be replaced to achieve effective anonymization\footnote{The demo will be available on GitHub page upon acceptance}.

\section{Evaluation}
\label{sec:eval}
Our evaluation protocol builds upon the VoicePrivacy Challenge (VPC) guidelines \cite{vpc2024, vpc2026}, pairing its downstream utility metrics with an uninformed attacker model for privacy validation. Speaker re-identification risks are quantified using the Equal Error Rate (EER) as the primary objective privacy metric. To evaluate downstream utility and naturalness, we measure two objective metrics: the Word Error Rate (WER) on an Automatic Speech Recognition (ASR) task to assess speech intelligibility, and the Unweighted Average Recall (UAR) on a Speech Emotion Recognition (SER) system to verify the preservation of emotional intent in the generated pseudo-speakers.
\subsection{Privacy}

For privacy evaluation via EER we employ an uninformed  attacker system. In this scenario, we employ an Automatic Speaker Verification (ASV) system based on the ECAPA-TDNN architecture \cite{desplanques2020ecapa} with 512 channels in its convolutional layers. This model is pre-trained on the VoxCeleb dataset \cite{Nagrani17voxceleb} using the SpeechBrain toolkit\footnote{\url{https://huggingface.co/speechbrain/spkrec-ecapa-voxceleb}}.

During evaluation, the attacker attempts to perform speaker re-identification by utilizing this model to extract fixed-dimensional speaker embeddings from both the original, unprotected enrollment utterances and the anonymized trial utterances. The cosine similarity score is then computed for each enrollment-trial pair. By varying the decision threshold across these similarity scores, the Equal Error Rate (EER) is calculated as the point where the False Acceptance Rate (FAR) and False Rejection Rate (FRR) converge, serving as our primary metric for identity concealment.

For the evaluation protocol, we randomly select 1,000 samples from the LibriTTS test-clean dataset\cite{zen2019libritts} to serve as user-provided data, denoted as trial utterances. We then process these through our tested systems to generate 1,000 anonymized utterances. A higher EER indicates greater privacy protection, with the maximum possible (random chance) EER bounded at 50\%.


\subsection{Utility}

To assess the anonymization system's ability to retain linguistic content, we employ an automatic speech recognition (ASR) system structured around the wav2vec 2.0 architecture \cite{baevski2020wav2vec}. Specifically, we utilize a speechbrain model\footnote{\url{https://huggingface.co/speechbrain/asr-wav2vec2-librispeech}} that leverages a \texttt{wav2vec2-large-960h-lv60-self} foundation model\footnote{\url{https://huggingface.co/facebook/wav2vec2-large-960h-lv60-self}} as a feature extractor, paired with a downstream Connectionist Temporal Classification (CTC) decoder. The entire joint network is fine-tuned on the 960-hour LibriSpeech \cite{Panayotov2015librispeech} corpus.

For evaluation, we decode the same 1,000 LibriTTS \cite{zen2019libritts}  trial utterances used in our speaker verification experiments. For every anonymized trial utterance, the ASR system generates a predicted word sequence. Intelligibility is then quantified via the Word Error Rate (WER) against the ground-truth transcripts. A lower WER indicates superior linguistic content preservation, verifying that the anonymization process does not corrupt semantic information.

Emotion preservation is evaluated using the Unweighted Average Recall (UAR). We employ a wav2vec2-based SER model\footnote{\url{https://huggingface.co/superb/wav2vec2-base-superb-er}} fine-tuned on the SUPERB benchmark \cite{yang2021superb}, which includes the IEMOCAP dataset \cite{busso2008iemocap} for training.
To evaluate the anonymized speech, we randomly sample 1,000 utterances from IEMOCAP, evenly distributed across four emotion classes: neutral, anger, happiness, and sadness. We synthesize each utterance through our anonymization system, perform emotion classification on the output. Emotion recognition performance is then quantified using the standard UAR metric: the sum of the class-wise recalls ($R_i$) divided by the total number of classes ($N_{class}$):
\begin{equation}
    UAR = \frac{\sum_{i=1}^{N_{class}} R_i}{N_{class}}.
\end{equation}
The recall $R_i$ for each class $i$ is computed as the number of true positives divided by the total number of samples in that class. A higher UAR indicates better emotion preservation.

\subsection{Speaker Diversity}
We evaluate the diversity and distributional alignment of the generated speaker embeddings using two metrics: Top-k Cosine Similarity and Maximum Mean Discrepancy (MMD) \cite{mmd}.

First, to assess speaker diversity, we compute the average Top-5 Cosine Similarity. For each of the 1,000 original speaker embeddings, we identify its 5 nearest neighbors with the highest cosine similarity within the dataset, and compute their mean cosine similarity. We repeat this process for the 1,000 embeddings generated under each experimental condition. A lower average cosine similarity indicates greater diversity among the speaker embeddings.

Second, we use MMD to measure the statistical divergence between the generated and original embedding distributions. Because the speaker embeddings are optimized in an angular margin space, we first apply $L_2$ normalization to all embeddings, mapping the Euclidean distances used by the Radial Basis Function (RBF) kernel\footnote{\url{https://scikit-learn.org/stable/modules/generated/sklearn.metrics.pairwise.rbf_kernel.html}} onto cosine similarity. Across all evaluated models, the intra-group MMD, which compares original embeddings to original and generated to generated, yielded consistently negligible values near zero. We therefore report only the cross-group metric (original vs. generated) in Table \ref{tab:speaker_anonymization_results} as the primary measure of distributional divergence. The higher the more difference of the distributions.

\section{Results}
\label{sec:res}

\subsection{Robustness against a semi-informed attacker}

\begin{table}[t]
\centering
\footnotesize 
\setlength{\tabcolsep}{3.5pt} 
\begin{tabular}{llrrr} 
\toprule
 &  & EER $\uparrow$ (\%) & WER $\downarrow$ (\%) & UAR $\uparrow$ (\%) \\ 
\midrule
\multicolumn{2}{l}{Orig} & 0.00 & 2.28 & 66.74 \\
\midrule
Cosyvoice & Recon & 7.88 & 3.99 & 42.37 \\ 
          & GMM       & 36.20 & \textbf{4.29} & 41.88 \\ 
          & NV      & 36.40 & \textbf{4.29} & \textbf{42.53} \\ 
\midrule
FACodec   & Recon & 6.26 & 4.50 & 36.58 \\ 
          & GMM       & \textbf{42.30} & 9.90 & 38.59 \\ 
          & NV      & 38.26 & 7.56 & 40.36 \\ 
\bottomrule
\end{tabular}
\caption{Privacy and utility evaluation results of original utterances (Orig), their reconstructions with two voice conversion systems (Recon) and anonymized utterances using the proposed NVAE pseudo speaker generator denoted as NV and the baseline GMM-based pseudo speaker generator. Best results marked in bold.}
\label{tab:evaluation}
\end{table}

\begin{figure*}[t]
    \centering
    \label{fig:layer_metrics}
    \includegraphics[width=1.0\linewidth]{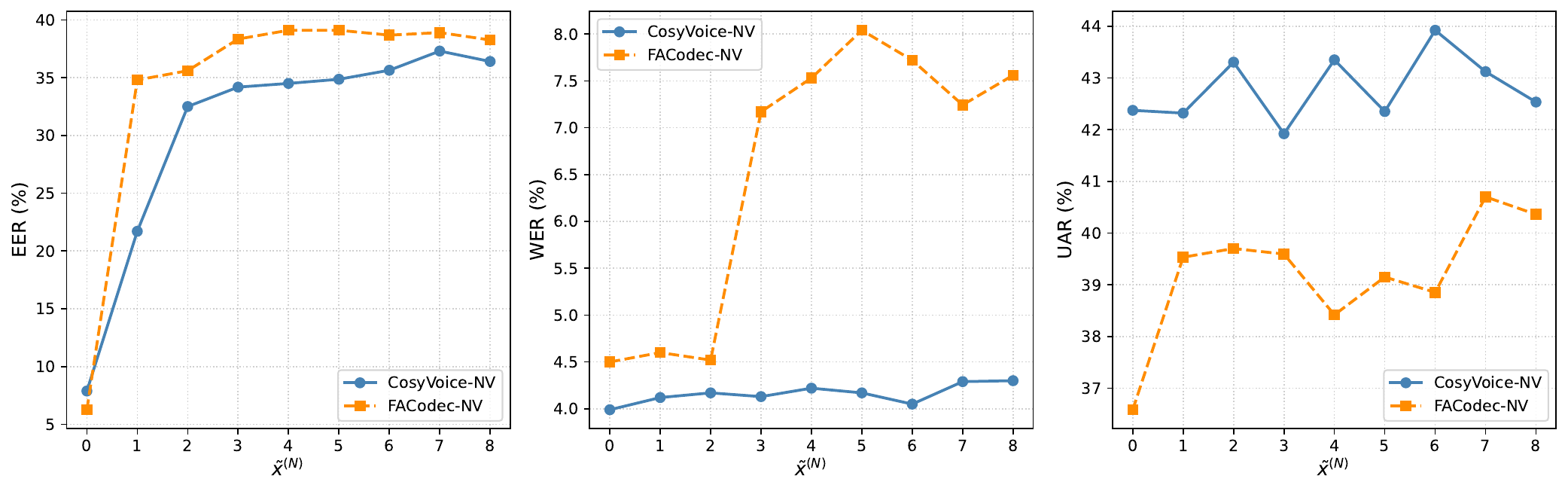}
    \caption{Effect of the number ($N$) of randomized groups of the NVAE latents on privacy (EER$\uparrow$) and utility (WER$\downarrow$ \& UAR$\uparrow$).}
\end{figure*}

Table \ref{tab:evaluation} reports the privacy and utility metrics for the unprocessed ground-truth audio ($Orig$), the reconstruction without anonymization ($Recon$), the GMM-based anonymization baseline, and the proposed NouveauVoice (NV) model.

The original recordings (Orig) provide perfect verifiability (0.00\% EER) and set the references for intelligibility (2.28\% WER) and emotion (66.74\% UAR). Passing audio through the reconstruction pipeline alone ($Recon$) raises the EER slightly and leaves intelligibility largely intact, but drops the UAR significantly (to 42.37\% for CosyVoice and 36.58\% for FACodec). This indicates that the large part of emotional information loss stems from the generation models rather than the anonymization process itself.

CosyVoice-NV matches or exceeds the CosyVoice-GMM baseline across all metrics. The two methods provide comparable privacy (36.40\% vs. 36.20\% EER) and identical intelligibility (4.29\% WER), while NVAE retains marginally more emotional expression (42.53\% vs. 41.88\% UAR) and performs on par with the $Recon$ baseline.

The comparison on FACodec models presents a clearer privacy-utility trade-off. FACodec-GMM shows higher anonymity (42.30\% vs. 38.26\% EER) but faces a pronounced cost to utility. The intelligibility degrades to 9.90\% WER (compared to 7.56\% for FACodec-NV), and its emotion preservation is lower too (38.59\% vs. 40.36\% UAR). Relatively, FACodec-NVAE exchanges 9.55\% decrease in EER for 23.64\% decrease in WER and 4.39\% increase in UAR, 
yielding a more favorable overall balance.


\subsection{Controlling anonymization strength}
To understand how the strength of anonymization can be controlled in NVAE layers, Figure \ref{fig:layer_metrics} reports performance as the NVAE hierarchy is progressively activated, as described in Section \ref{sec:exp2}.

For CosyVoice-NV, privacy increases smoothly and monotonically with the number of active layers, rising from 7.88\% EER at $\tilde{x}^{(0)}$ to 36.40\% at $\tilde{x}^{(8)}$, while utility remains mostly flat throughout. The majority of privacy gain is realized by the first three layers. The EER jumps by 13.8 and 10.8 points at $\tilde{x}^{(1)}$ and $\tilde{x}^{(2)}$, after which randomizing additional layers brings diminishing improvements. CosyVoice-NV therefore enables a flexible, controllable tradeoff in which speaker anonymity can be strengthened at almost no cost to intelligibility or emotional expressiveness.

FACodec-NV shows an anonymization pattern similar to CosyVoice-NV, where the privacy plateaus at $\tilde{x}^{(2)}$. However, intelligibility is preserved only up to $\tilde{x}^{(2)}$ (4.52\% WER) before degrading sharply and exceeding 7\% from $\tilde{x}^{(3)}$ and onward. This highlights a tradeoff between anonymization strength and intelligibility performance. This reveals $\tilde{x}^{(2)}$ as the optimal setting for FACodec, as it captures most of the available privacy (35.60\% EER) while keeping intelligibility near the reconstruction baseline (4.50\% WER).

\subsection{Diversity and Distributional Alignment of Generated Embeddings}
Table~\ref{tab:speaker_anonymization_results} demonstrates that the proposed NVAE models achieve substantially greater speaker embedding diversity than the GMM baselines. Across both FACodec-NV and CosyVoice2-NV, deeper NVAE layers shows progressively lower Top-5 Cosine Similarity, with CosyVoice2-NV reaching as low as 0.40 compared to the original 0.75. 

By comparison, the GMM baselines keep the speakers very similar to the original and match the overall data distribution better, as shown by their lower MMD scores. However, they don't provide much diversity gain. This result shows that generating new, varied voices inherently pushes the embeddings further away from the original distribution.

\begin{table}[ht]
\centering
\caption{Comparison of Speaker Similarity and Diversity Metrics Across Varied NVAE Layers and GMM Baselines}
\label{tab:speaker_anonymization_results}
\begin{tabular}{lcc}
\toprule
\textbf{Configuration} & \textbf{Top-5 Cosine Similarity} & \textbf{MMD} \\
& (Mean $\pm$ SD) & \\
\midrule
\textit{FACodec-NV} & \multicolumn{2}{l}{[Orig Similarity: $0.84 \pm 0.06$]} \\
\cmidrule(lr){1-3}
\quad Layer 0 & $0.80 \pm 0.07$ & 0.0053 \\
\quad Layer 1 & $0.68 \pm 0.07$ & 0.0680 \\
\quad Layer 2 & $0.68 \pm 0.07$ & 0.0642 \\
\quad Layer 3 & $0.68 \pm 0.07$ & 0.1086 \\
\quad Layer 4 & $0.68 \pm 0.07$ & 0.1086 \\
\quad Layer 6 & $0.67 \pm 0.07$ & 0.1064 \\
\quad Layer 8 & $0.66 \pm 0.06$ & 0.1022 \\
\quad GMM Baseline & $0.80 \pm 0.07$ & 0.0153 \\
\midrule
\textit{CosyVoice2-NV} & \multicolumn{2}{l}{[Orig Similarity: $0.75 \pm 0.05$]} \\
\cmidrule(lr){1-3}
\quad Layer 0 & $0.77 \pm 0.07$ & 0.0043 \\
\quad Layer 1 & $0.71 \pm 0.04$ & 0.0130 \\
\quad Layer 2 & $0.49 \pm 0.03$ & 0.1318 \\
\quad Layer 3 & $0.46 \pm 0.03$ & 0.1555 \\
\quad Layer 4 & $0.44 \pm 0.03$ & 0.1436 \\
\quad Layer 6 & $0.40 \pm 0.03$ & 0.1603 \\
\quad Layer 8 & $0.42 \pm 0.03$ & 0.1690 \\
\quad GMM Baseline & $0.74 \pm 0.06$ & 0.0722 \\
\bottomrule
\end{tabular}
\end{table}

\section{Conclusion and Future Work}
\label{sec:conc}
We propose NouveauVoice, a novel pseudo-speaker generation framework based on a Hierarchical Deep Variational Autoencoder. Designed as a standalone plug-in module, it integrates seamlessly with state-of-the-art architectures  with minimal training. The framework generates highly expressive, diverse pseudo-speaker embeddings for use in the anonymization process. Evaluation results demonstrate that NouveauVoice achieves strong identity protection while successfully preserving speech intelligibility and utility. Furthermore, our ablation study shows that by controlling the number of layers in the NVAE, one can achieve different levels of anonymization, offering a flexible privacy-utility tradeoff. Future work includes developing an attribute-controlled (age, gender, accent) speaker generation model using guided latent space sampling.

\section{Acknowledgments}
The authors acknowledge the use of generative AI tools (such as Large Language Models) solely for English language editing, grammatical correction, and structural rephrasing of the manuscript during the writing process. All technical content, experimental designs, and conclusions remain entirely the responsibility of the authors.

\newpage
\bibliographystyle{IEEEbib}
\bibliography{refs}

\end{document}